\documentclass[twocolumn]{aastex62}

\newcommand\apjcls{1}
\newcommand\aastexcls{2}
\newcommand\othercls{3}

\newcommand\papercls{\aastexcls}
\if\papercls \apjcls
\usepackage{apjfonts}
\else\if\papercls \othercls
\usepackage{epsfig}
\usepackage{margin}
\usepackage{times}
\fi\fi
\usepackage{ifthen}
\usepackage{natbib}
\usepackage{bm}
\usepackage{amssymb, amsmath}
\usepackage{appendix}
\usepackage{etoolbox}
\usepackage[T1]{fontenc}
\usepackage{paralist}
\usepackage{newtxtext,newtxmath}

\if\papercls \apjcls
\newcommand\aas{\ref@jnl{AAS Meeting Abstracts}}
\newcommand\dps{\ref@jnl{AAS/DPS Meeting Abstracts}}
\newcommand\maps{\ref@jnl{MAPS}}
\else\if\papercls \othercls
\usepackage{astjnlabbrev-jh}
\fi\fi

\if\papercls \aastexcls
\hypersetup{citecolor=blue, 
            linkcolor=blue, 
            menucolor=blue, 
            urlcolor=blue}  
\else
\usepackage[
bookmarks=true,           
bookmarksnumbered=true,   
colorlinks=true,          
citecolor=blue,           
linkcolor=blue,           
menucolor=blue,           
urlcolor=blue,            
linkbordercolor={0 0 1},  
pdfborder={0 0 1},
frenchlinks=true]{hyperref}
\fi
\if\papercls \othercls

\else

\fi

\providecommand{\adsurl}[1]{\href{#1}{ADS}}

\makeatletter
\patchcmd{\NAT@citex}
  {\@citea\NAT@hyper@{%
     \NAT@nmfmt{\NAT@nm}%
     \hyper@natlinkbreak{\NAT@aysep\NAT@spacechar}{\@citeb\@extra@b@citeb}%
     \NAT@date}}
  {\@citea\NAT@nmfmt{\NAT@nm}%
   \NAT@aysep\NAT@spacechar\NAT@hyper@{\NAT@date}}{}{}

\patchcmd{\NAT@citex}
  {\@citea\NAT@hyper@{%
     \NAT@nmfmt{\NAT@nm}%
     \hyper@natlinkbreak{\NAT@spacechar\NAT@@open\if*#1*\else#1\NAT@spacechar\fi}%
       {\@citeb\@extra@b@citeb}%
     \NAT@date}}
  {\@citea\NAT@nmfmt{\NAT@nm}%
   \NAT@spacechar\NAT@@open\if*#1*\else#1\NAT@spacechar\fi\NAT@hyper@{\NAT@date}}
  {}{}
\makeatother

\makeatletter
\DeclareRobustCommand{\lowcase}[1]{\@lowcase#1\@nil}
\def\@lowcase#1\@nil{\if\relax#1\relax\else\MakeLowercase{#1}\fi}
\makeatother

\DeclareSymbolFont{UPM}{U}{eur}{m}{n}
\DeclareMathSymbol{\umu}{0}{UPM}{"16}
\let\oldumu=\umu
\renewcommand\umu{\ifmmode\oldumu\else\math{\oldumu}\fi}

\if\papercls \othercls

\else

\fi

\let\oldsim=\sim
\renewcommand\sim{\ifmmode\oldsim\else\math{\oldsim}\fi}
\let\oldpm=\pm
\renewcommand\pm{\ifmmode\oldpm\else\math{\oldpm}\fi}
\newcommand\by{\ifmmode\times\else\math{\times}\fi}

\newbox{\wdbox}
\renewcommand\c{\setbox\wdbox=\hbox{,}\hspace{\wd\wdbox}}
\renewcommand\i{\setbox\wdbox=\hbox{i}\hspace{\wd\wdbox}}

\newcount\timect
\newcount\hourct
\newcount\minct
\newcommand\now{\timect=\time \divide\timect by 60
         \hourct=\timect Cltiply\hourct by 60
         \minct=\time \advance\minct by -\hourct
         \number\timect:\ifnum \minct < 10 0\fi\number\minct}

\catcode`@=11

\newcommand\comment[1]{}

\newcommand\commenton{\catcode`\%=14}

\renewcommand\math[1]{$#1$}
\newcommand\mathshifton{\catcode`\$=3}

\let\atab=&
\newcommand\atabon{\catcode`\&=4}

\let\oldmsp=\sp
\let\oldmsb=\sb
\def\sp#1{\ifmmode
           \oldmsp{#1}%
         \else\strut\raise.85ex\hbox{\scriptsize #1}\fi}
\def\sb#1{\ifmmode
           \oldmsb{#1}%
         \else\strut\raise-.54ex\hbox{\scriptsize #1}\fi}
\newbox\@sp
\newbox\@sb
\def\sbp#1#2{\ifmmode%
           \oldmsb{#1}\oldmsp{#2}%
         \else
           \setbox\@sb=\hbox{\sb{#1}}%
           \setbox\@sp=\hbox{\sp{#2}}%
           \rlap{\copy\@sb}\copy\@sp
           \ifdim \wd\@sb >\wd\@sp
             \hskip -\wd\@sp \hskip \wd\@sb
           \fi
        \fi}
\def\msp#1{\ifmmode
           \oldmsp{#1}
         \else \math{\oldmsp{#1}}\fi}
\def\msb#1{\ifmmode
           \oldmsb{#1}
         \else \math{\oldmsb{#1}}\fi}

\def\supon{\catcode`\^=7}

\def\subon{\catcode`\_=8}

\def\supsubon{\supon \subon}

\newcommand\actcharon{\catcode`\~=13}

\newcommand\paramon{\catcode`\#=6}

\comment{And now to turn us totally on and off...}

\newcommand\reservedcharson{ \commenton  \mathshifton  \atabon  \supsubon 
                             \actcharon  \paramon}

\catcode`@=12
\reservedcharson

\if\papercls \apjcls

\else

\fi

\newcommand\chisq{\ifmmode{\chi\sp{2}}\else\math{\chi\sp{2}}\fi}
\newcommand\redchisq{\ifmmode{ \chi\sp{2}\sb{\rm red}}
                    \else\math{\chi\sp{2}\sb{\rm red}}\fi}
\newcommand\Teq{\ifmmode{T\sb{\rm eq}}\else$T$\sb{eq}\fi}
\newcommand\mjup{\ifmmode{M\sb{\rm Jup}}\else$M$\sb{Jup}\fi}
\newcommand\rjup{\ifmmode{R\sb{\rm Jup}}\else$R$\sb{Jup}\fi}
\newcommand\msun{\ifmmode{M\sb{\odot}}\else$M\sb{\odot}$\fi}
\newcommand\rsun{\ifmmode{R\sb{\odot}}\else$R\sb{\odot}$\fi}
\newcommand\mearth{\ifmmode{M\sb{\oplus}}\else$M\sb{\oplus}$\fi}
\newcommand\rearth{\ifmmode{R\sb{\oplus}}\else$R\sb{\oplus}$\fi}

\newcommand\Pran{\ensuremath{\mathrm{Pr}}}

\renewcommand{\bm}[1]{{\mbox{{\boldmath$#1$}}}}	

\usepackage{CJK}

\begin{document}

\title{3D Simulations of Convective Entrainment in Gas Giants: Rotation and Decreasing Luminosity as Barriers to Mixing}

\author{\begin{CJK*}{UTF8}{gbsn}Shu Zhang (张舒)\end{CJK*}}

\affiliation{\rm Department of Physics and Trottier Space Institute, McGill University, Montreal, QC H3A 2T8, Canada}

\author{J.~R.~Fuentes}
\affiliation{\rm TAPIR, California Institute of Technology, Pasadena, CA 91125, USA}

\author{Andrew Cumming}
\affiliation{\rm Department of Physics and Trottier Space Institute, McGill University, Montreal, QC H3A 2T8, Canada}

\begin{abstract}
Observations from Juno and Cassini suggest that Jupiter and Saturn may possess fuzzy cores -- central regions where the abundance of heavy elements varies smoothly with depth. Such gradients pose a longstanding puzzle for models of planetary evolution and formation, which predict that vigorous convection would homogenize the interior of gas giants within the first $\sim 10^6$--$10^8~\mathrm{years}$ of cooling. Previous 3D simulations and analytic predictions for the propagation of a convection zone into a stable region have demonstrated that the rapid rotation of gas giants can significantly slow convective mixing, but not enough to stop it. Another piece of the puzzle is luminosity. Gas giants cool as they age, and with that comes a declining heat flux over time. Recent ideas suggest that when this declining luminosity is combined with rotational effects, convection may stall. We explore this possibility using 3D hydrodynamic simulations that include both rotation and a surface cooling flux that decreases as $1/t$. Our results demonstrate that, even without rotation, a declining luminosity can suppress mixing sufficiently to preserve an initial compositional gradient in the deep interior of gas giants. If confirmed by more realistic simulations, this may help to explain the long-term survival of fuzzy cores.
\end{abstract}


\section{Introduction}
\label{sec:Introduction}

Modern interior models of Jupiter and Saturn, which fit observational constraints from gravitational fields (for both planets) and ring seismology (for Saturn), suggest inhomogeneous interiors, some of them predicting extended composition gradients \citep[e.g.,][]{Fuller2014, Wahl2017,Debras2019,Mankovich2021, Nettelmann2021, Militzer2022, Miguel2022, Idini2022, Helled2022, Dewberry2023, Howard_2023}. Currently, no consensus on a formation pathway to these current-day structures has emerged \citep{Liu2019,Muller2020,Stevenson2022,Bodenheimer2025}. These findings caused significant interest in gaining a better understanding of the mixing processes inside giant planets, particularly how convection--driven by cooling from the planet's surface--may lead to the dispersal and/or homogenization of their cores.

In the past few years, several studies have explored this topic using 1D thermal evolution models \citep[e.g.,][]{Muller2020,Knierim2024,Arevalo2025}. A key result from these works is that if a planet forms hot--with a high internal entropy--vigorous convective mixing can take place on short timescales ($\sim$ 1 Myr), in some cases nearly homogenizing the entire planet's interior. These simulations, however, lack a detailed model of how composition and energy are transported at
convective boundaries. In an attempt to better model convective boundary mixing, \cite{Fuentes2020} used 2D hydrodynamic simulations of a local Cartesian box to investigate the propagation of a convection zone into a stably-stratified region by a composition gradient. Their experiments, which aimed to model convective mixing between the planet's core and the envelope, showed that composition gradients are ineffective barriers against moving convective zones. Similar results were found by \cite{Anders2022} in the context of core-convection in stars.

Another way the planet's interior could avoid complete mixing is through the formation of a semiconvective staircase beneath the outer convective envelope \citep[e.g.,][]{Radko2003,Leconte2012,Leconte2013}. In this scenario, the competition between thermal and compositional gradients in the deep interior triggers double-diffusive instabilities. The nonlinear outcome is a remarkable form of self-organization, a series of convective layers separated by thin, stably stratified interfaces \citep[see, e.g.,][for a review]{Garaud2018}. However, this mechanism has been challenged by hydrodynamical simulations, showing that staircases rapidly mix into a single convective layer \citep[e.g.,][]{Wood2013,Moll2017,Fuentes2022,Tulekeyev2024}.

Rapid rotation, inherently present in giant planets, is well-known to reduce the efficiency of convective heat transport when the Rossby number $\mathrm{Ro}$ --- the ratio of the rotation period to the convective turnover time --- is much smaller than one \citep[e.g.,][]{Julien1996_2}. Building on this, \cite{Fuentes2023} used classical scaling arguments for convective velocities in rapidly rotating flows \citep[e.g.,][]{Stevenson1979,Barker2014,Aurnou2020} together with 3D simulations in boxes to revisit the problem of a convection zone advancing into a stably stratified region. They found that for typical values of the Rossby number in gas giants ($\mathrm{Ro} \sim 10^{-6}$; \citealt{Guillot_2004}), the mixing time of primordial composition gradients can be delayed by more than two orders of magnitude, suggesting that rotation plays an important role to preserve inhomogeneous interiors. Along similar lines, \citet{Fuentes2024}, using 3D local Cartesian simulations, and \citet{Fuentes2025}, using simulations in the full sphere, demonstrated that rotation also slows the merger of semiconvective layers. Yet, in all cases, the simulations ultimately evolved toward a fully convective state, i.e., mixing is slowed, not stopped.

Seeking a mechanism to fully suppress convective mixing, \cite{Hindman2023} made an analytic model for the propagation of a convective interface. They suggested that convection in a gas giant can stall before fully mixing the interior, not because of a strong stabilizing composition gradient, but because the system runs out of energy. As the luminosity declines, convective motions weaken. This, in combination with rotation further suppressing the ability to entrain deeper material, causes the growing convective zone to stall, leaving behind a partially unmixed interior. While promising, this theory has yet to be tested with hydrodynamical simulations. In this paper, we aim to test these ideas. In Section~\ref{sec:scalings} we summarize the analytic predictions of \cite{Hindman2023}. In Section~\ref{sec:numerics} we present our numerical experiments and results. We conclude in Section~\ref{sec:discussion} with a summary and discussion of how our results apply to gas giants.

\section{Mixing-Length Scalings and Convective Entrainment}\label{sec:scalings}

In this section, we summarize the mixing-length scalings for convective velocities in both nonrotating and rotating flows, and discuss how these scalings are applied to model the propagation of a convection zone into a stably stratified compositional gradient.

When the boundary between the convection zone and the adjacent stable region is thin, energetic arguments predict the speed at which the convection zone propagates.  The entrainment hypothesis \citep[e.g.,][]{Linden1975,Fernando1987,Molemaker1997} states that the gravitational potential energy required to mix denser fluid with the overlying light fluid within the convection zone must come from the kinetic energy of the convective motions; explicitly, the rate of potential energy increase due to the upward transport
of heavy material is proportional to the vertical kinetic energy flux carried by the convection. Such a hypothesis leads to a differential equation for the depth of the convection zone, $h$,
\begin{equation} \label{eqn:hypothesis}
    \frac{1}{2}g h \, \Delta\rho \frac{dh}{dt} \approx \frac{\gamma}{2}\rho_0 U^3_{c}\,,
\end{equation}
where  $g$ is the acceleration due to gravity, $\rho_0$ is the mass density within the convection zone, $\Delta\rho$ is the density jump across the thin interface, $U_c$ is a typical convective flow speed, and the proportionality constant $\gamma$ (known as the mixing efficiency) is of order unity for turbulent flows \citep{Fuentes2020}. We emphasize that $\Delta \rho$ arises because composition in the convection zone is well mixed, creating a contrast in density with the underlying region which has a smooth composition gradient.

This entrainment equation can be solved for the thickness of the convection zone $h$ as a function of time, given prescriptions for the density jump across the convective interface $\Delta \rho$ and convective speed $U_c$. In the non-rotating and rapidly-rotating limits, typical convective speeds ($U_{\rm NR}$ and $U_{\rm R}$, respectively) can be estimated using mixing-length theory and turbulent, diffusion-free scaling laws for convective heat transfer \citep[e.g.,][] {Barker2014, Aurnou2020, Fuentes2023}
\begin{equation}
U_{\mathrm{NR}}\approx \left(\dfrac{\alpha g F_T h}{\rho_0 c_P}\right)^{1/3}\,,\hspace{0.25cm} U_{\mathrm{R}}
\approx \left(\dfrac{\alpha g F_T}{\rho_0 c_P}\right)^{2/5}\left(\dfrac{h}{2\Omega}\right)^{1/5}~.\label{eq:u_conv}
\end{equation}
In these expressions, $F_T$ is the  heat flux applied at the top boundary, $\alpha = -(\partial \ln \rho/\partial T)_C$ is the coefficient of thermal expansion at fixed composition, $c_P$ is the specific heat capacity at constant pressure, and $\Omega$ is the rotation rate.

On other hand, the density jump across the interface is determined by the corresponding jumps in composition and temperature, 

\begin{equation}
    \Delta \rho = \rho_0(\beta \Delta C - \alpha \Delta T)~,
\end{equation}
where $\beta = (\partial \ln\rho/\partial C)_{T}$ is the coefficient of compositional contraction at fixed temperature, and both $\Delta C$ and $\Delta T$ can be obtained by applying conservation of mass and energy within a layer of size $h$ before and after convective mixing. Since a sharp stably-stratified interface is dominated by the compositional jump across\footnote{For example, \cite{Fuentes2020} found that the $\Delta T$ term contributes a $\approx 20$--$30$\% correction to $\Delta \rho$ at low Prandtl number (see their Fig.~5).}, \cite{Hindman2023} assumed $\Delta \rho \approx \rho_0\beta\Delta C$. For a constant driving heat flux and an initially linear density stratification, integration of Equations \eqref{eqn:hypothesis}--\eqref{eq:u_conv} in the limit of no rotation and neglecting heat diffusion across the interface yields the well-known result that the convective layer grows with a square root of time dependence
\begin{equation}\label{eq:h_FF_no_rot}
h_{\rm NR}(t) = \left[h_0^2 + 4 \gamma H^2 \, Nt\right]^{1/2}~,
\end{equation}
where $h_0$ is an integration constant, $N^2 = g\beta |dC_0/dz|$ is the initial buoyancy frequency and $H = (g\alpha F_T/\rho c_P N^3)^{1/2}$ is an overshooting length scale that characterizes the depth to which convective plumes penetrate locally into the stable region \citep{Molemaker1997,Fuentes2020}. However, in the presence of rapid rotation, the growth of the convective layer is slowed, following a power law with an index of 5/12 \citep{Fuentes2023}
\begin{equation}\label{eq:h_FF_rot}
 h_{\rm R}(t) \approx \left[h_0^{12/5} + \frac{24 \gamma}{5} H^{12/5} \!\left(\frac{N}{2\Omega}\right)^{\!\!3/5} \!\!(Nt) \right]^{5/12}~.
\end{equation}

Adapting the model of \citet{Fuentes2023} to include a more realistic, time-dependent boundary condition for the heat flux, $F_T = F_0(t_0/t)$, \citet{Hindman2023} developed an analytic model and showed that in rapidly rotating flows, the growth of the convective layer eventually stalls, whereas in the absence of rotation, the layer continues to deepen over time

\begin{gather}
    h_{\rm NR}(t) \approx \left[h_0^2 + 4\gamma H_0^2 \, (Nt_0)\ln(t/t_0)\right]^{1/2}~, \label{eq:h_DF_no_rot} \\
h_{\rm R}(t) \approx \left[h_0^{12/5} + 24\gamma H_0^{12/5} (Nt_0)\left(\frac{N}{2\Omega}\right)^{\!\!3/5}\!\! \left(1- \frac{t_0^{1/5}}{t^{1/5}}\right)\right]^{5/12}~, \label{eq:h_DF_rot}
\end{gather}
where $H_0 = (g\alpha F_0/\rho c_P N^3)^{1/2}$. While the predictions in Equations \eqref{eq:h_FF_no_rot} and \eqref{eq:h_FF_rot} have been investigated and validated through 3D numerical simulations \citep{Fuentes2023}, the predictions corresponding to the time-dependent flux in Equations \eqref{eq:h_DF_no_rot} and \eqref{eq:h_DF_rot} have not yet been verified by numerical simulations. The stalled solution is particularly relevant to the internal structure of gas giants, as it implies the preservation of primordial composition gradients beneath the convective front. In this work, we verify these predictions using 3D hydrodynamical simulations.

\section{Numerical Simulations}\label{sec:numerics}
\begin{figure}
    \centering
    \includegraphics[width=\columnwidth]{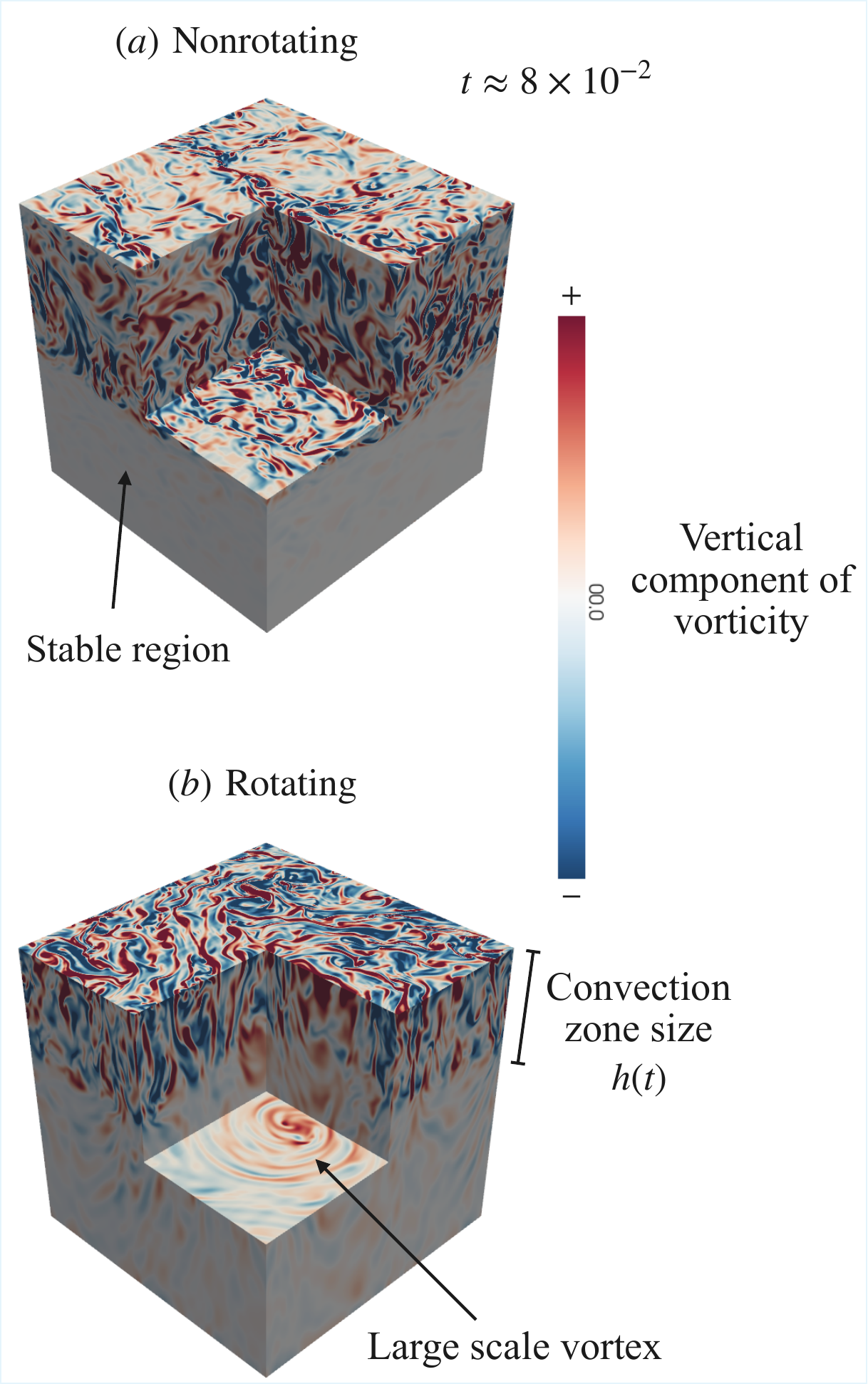}
    \caption{3D snapshots of the $z$ component of the vorticity $\bm{\omega} = \nabla \times \bm{u}$ for the nonrotating case (panel a) and rotating case (panel b), both cooling with a decreasing flux from the top boundary. For both cases, the snapshots are shown at the same time, when the heat flux that drives convection has decreased to $10\%$ of its initial value. The Rossby number of the convective flow in the rotating case is $\mathrm{Ro} \approx 0.01$, computed using the averaged flow velocity in the convection zone, and thickness of the convection zone. At $t=8\times 10^{-2}$, the convection zone depth in units of the box size $H$ is $h = 0.37$ for the rotating case, and $h = 0.56$ in the non-rotating case.}
    \label{fig_3d_sim}
\end{figure}

\begin{figure*}
    \centering
    \includegraphics[width=0.9\textwidth]{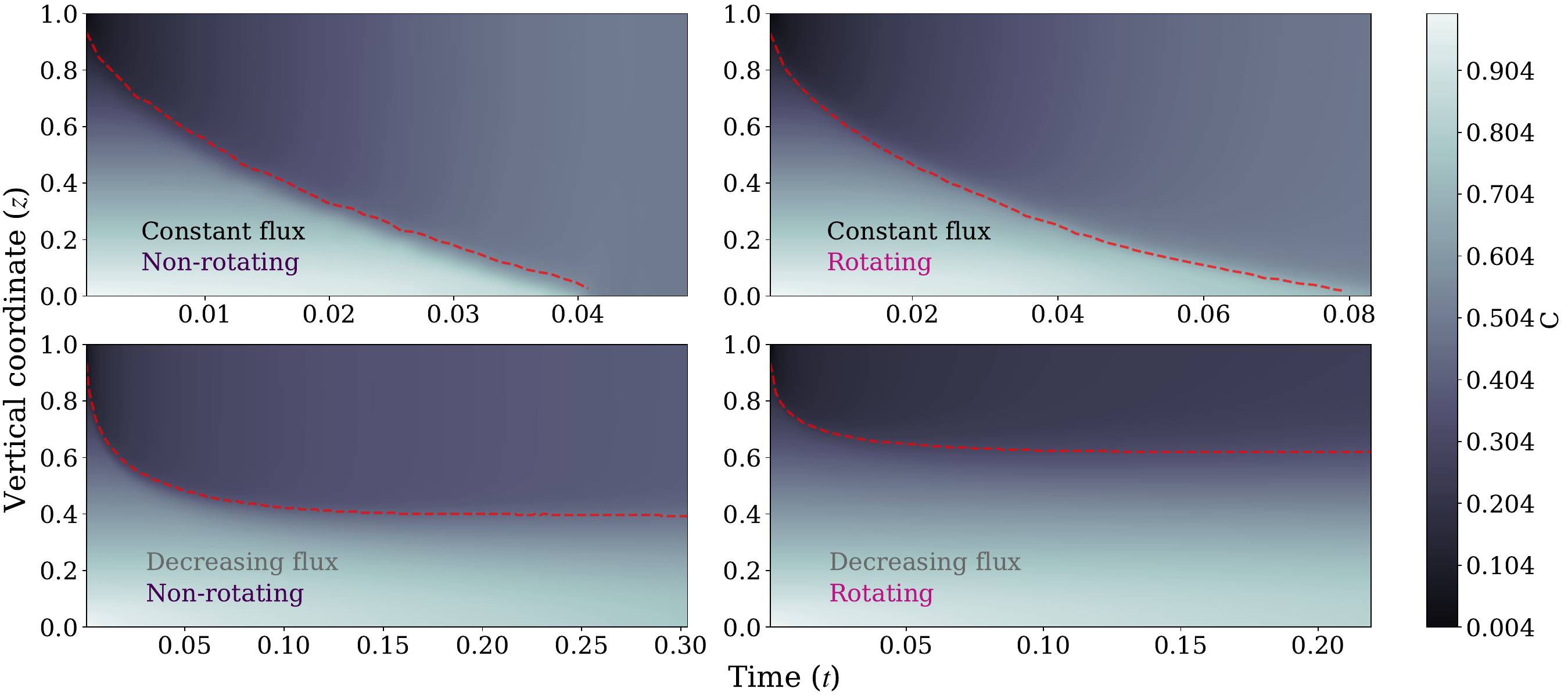}
    \caption{Color maps showing the evolution of the horizontally-averaged composition over time. In each case, the red dashed line shows the measured location of the convection zone boundary.}
    \label{fig:composition_maps}
\end{figure*}

We present four 3D simulations of penetrative convection, both with and without rotation, in a plane-parallel domain with equal horizontal and vertical extent $H$. We follow the same numerical setup and use the same code as described in \cite{Fuentes2023}. The lateral boundaries are periodic, while the top and bottom are impermeable and stress-free, with zero compositional flux. The fluid is initially stabilized by a linear compositional gradient. Cooling is imposed at the top boundary by fixing the thermal gradient with a negative heat flux, either held constant at $F_0$, or allowed to decay with time as $F_0~(t_\star/(t_\star + t))$, where we introduced $t_\star \sim 5\times 10^{-3} \ll t$ to avoid divergence of the flux at $t=0$. In both cases, the effect is to destabilize the upper layers until convection sets in. As the system evolves, convection progressively erodes the compositional gradient underneath, establishing a well-mixed region that deepens with time. We investigate how the growth rate of convective zone changes by with the imposed heat flux and rotation.

In our simulations, as in the predictions above, we assume that fluid motions occur on scales much smaller than a density scale height. This justifies neglecting curvature effects and adopting the Boussinesq approximation \citep{Spiegel_Veronis_1960}. Although the density in giant planets can vary by many orders of magnitude, this approximation fully captures nonlinear mixing near the convective boundary, which is the focus of our study. In the rotating cases, we consider uniform rotation throughout the fluid, and assume that gravity and rotation are aligned and point in the z direction, $\bm{\Omega} = \Omega \hat{\bm{z}}$, and $\bm{g} = -g\hat{\bm{z}}$, i.e., the simulation is at a polar latitude. The density of the fluid depends on both composition and temperature fluctuations. In what follows, all results are presented in dimensionless form, with time normalized to the thermal diffusion time across the box, $\tau_{T} = H^2/\kappa_T$, where $\kappa_T$ is the thermal diffusivity, and length normalized to the box height $H$.  For more details on the numerical setup, nondimensionalization, and code, we refer the reader to the Appendix.

Figure~\ref{fig_3d_sim} shows 3D snapshots of the $z$ component of the fluid vorticity, $\hat{\bm{z}}\cdot (\nabla \times \bm{u})$. Since both non-rotating and rotating cases with fixed-flux, boundary-driven convection zones have already been presented in \cite{Fuentes2023}, we focus here on new simulations where convection is driven by a gradually declining heat flux. Snapshots are shown for both the non-rotating (panel a) and rotating (panel b) cases at the same time, $t\approx 8\times 10^{-2}$, when the flux heat has decreased to about $10\%$ of its initial value. As expected, in the absence of rotation, the convection zone has entrained and mixed more fluid than in the rotating case, since the kinetic energy flux available for mixing is larger. The flow is also isotropic, with length scales that are approximately equal in all directions. In contrast, in the rotating case, the convective flow is anisotropic, forming vortices and columns whose horizontal length scales are much smaller than the depth of the convection zone. 

Figure~\ref{fig:composition_maps} shows the temporal evolution of the horizontally-averaged composition. The dashed curves show the extent of the convection zone as measured from density profiles. For constant heat flux, convection mixes the entire box as expected. When the flux decreases with time, the convection zone stalls even without rotation. This departs from the predictions in Section~\ref{sec:scalings}, where stalling was expected only in the rotating case. 

To test the analytic predictions, we first verify that both scalings for the convective velocity in Equation~\eqref{eq:u_conv} are consistent with our numerical experiments. To do so, we compute the average vertical component of the rms velocity across the convection zone.

In the rotating cases, a large-scale vortex develops over time, enhancing the horizontal velocities, often by an order of magnitude compared to the vertical flow. The analytical model does not capture this vortex, including only the vertical velocity.  However, we find that it reproduces the numerical results well even in the presence of the vortex, suggesting that the vortex does not affect the entrainment process and that the entrainment is set by the strength of the vertical velocity.

\begin{figure*}
    \centering
    \includegraphics[width=\textwidth]{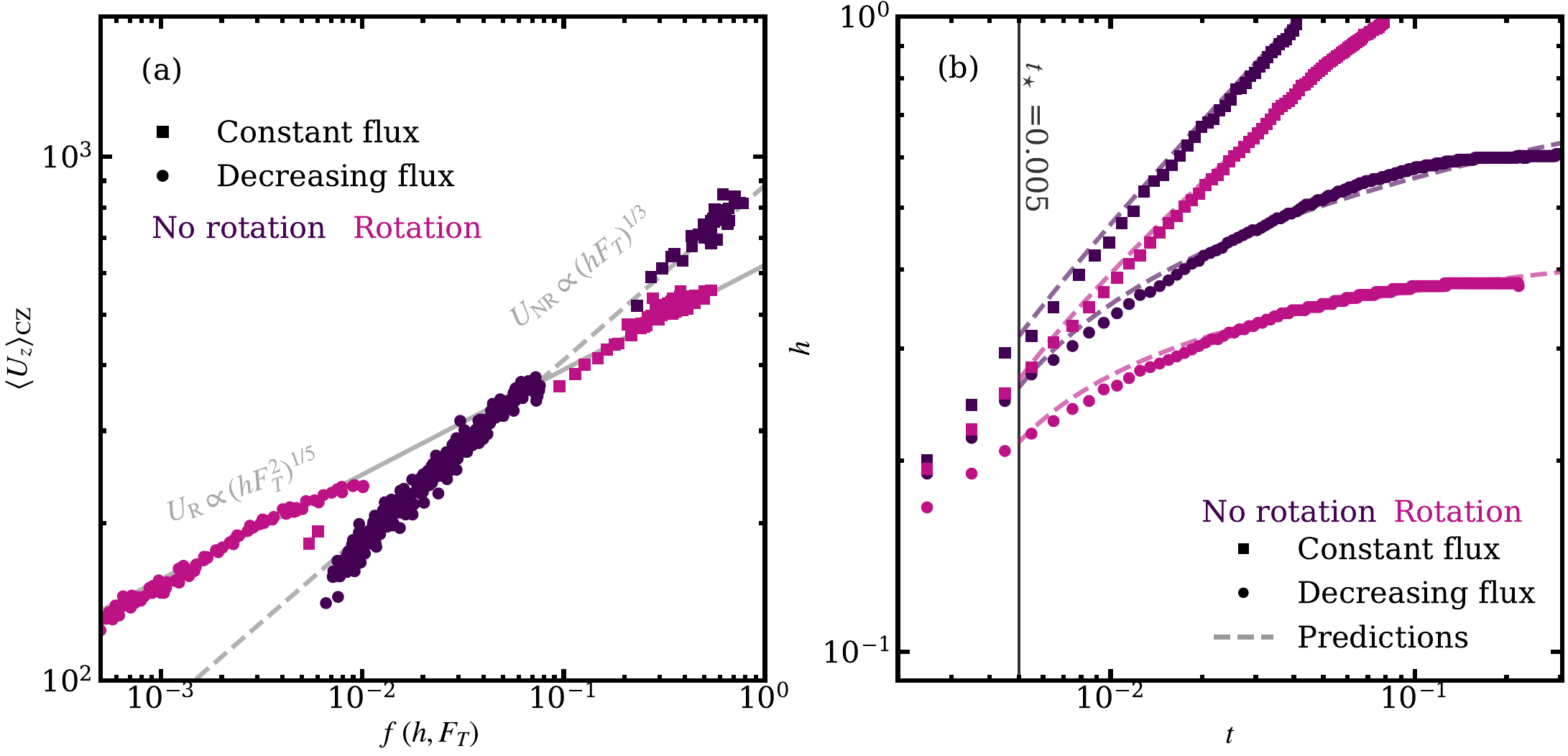}
    \caption{Panel (a): Vertical component of the rms flow velocity, averaged over the convection zone, as a function of convection zone thickness $h$ and imposed heat flux $F_T$. The quantity plotted on the x-axis is $f=hF_T$ for the non-rotating case, and $f=hF_T^2$ for the rotating case. The dashed and solid lines represent the theoretical predictions from Equation~\eqref{eq:u_conv} (with prefactors of 0.7 and 0.95 respectively to obtain agreement with the measured values). Panel (b): Thickness of the convection zone, $h$, as a function of time, $t$, for all our simulations. The vertical line marks $t_{\star}=0.005$, the time at which the cooling flux at the top boundary begins to decrease with time. The dashed lines correspond to the the analytic predictions in Equations~\eqref{eq:h_FF_no_rot}--\eqref{eq:h_DF_rot}, with best-fit values $\gamma=0.58,1.9,0.41,0.022$ from top to bottom.}    
    \label{fig:fits}
\end{figure*}

As shown in Figure~\ref{fig:fits}(a), the scaling of the convective velocity with the heat flux and the thickness of the convective zone is well supported by our simulations, except at early times, when the convection zone is still shallow and not fully developed.

Figure~\ref{fig:fits}(b) shows the measured size of the convection zone compared with the analytic predictions. 
Overall, the simulations are consistent with the theory in that, for rotating cases, the rate of entrainment and the growth of the convective layer are significantly reduced compared to the nonrotating cases. For the cases with constant flux, there is excellent agreement with Equations~\eqref{eq:h_FF_no_rot}--\eqref{eq:h_FF_rot}. In contrast, for the cases with a decreasing flux, we observe noticeable deviations from Equations~\eqref{eq:h_DF_no_rot} and \eqref{eq:h_DF_rot}, specifically the measured $h(t)$ curves become much flatter than predicted at late times. Preliminary analyses suggest that these discrepancies arise from a time-dependent mixing efficiency in the entrainment equation. In cases with declining flux, the efficiency decreases asymptotically to 0, whereas it remains approximately constant and of order unity in cases with constant flux. A complete characterization of the mixing efficiency requires a thorough analysis of the heat and composition transport across the interface, which is beyond the scope of this paper. Nevertheless, we emphasize that the most important result, the stall of the convection zone, is confirmed by our hydrodynamical simulations. Furthermore, even the nonrotating case seems to stall.

\section{Discussion and Conclusion}\label{sec:discussion}

In this work, we explored how the deep interiors of gas giants may resist convective mixing when the planet cools over time with a decreasing luminosity. Using 3D hydrodynamical simulations, we tested recent analytical arguments by \citet{Hindman2023} (see also Equations~\ref{eq:h_FF_no_rot}--\ref{eq:h_DF_rot}), which suggest that convection can stall at a finite depth because rotation and a declining heat flux act together to inhibit overturning motions. In this picture, the convective envelope fails to reach the core, potentially preserving a primordial interior that remains largely untouched since formation.

As shown in Figure~\ref{fig:fits}, there is excellent agreement between the simulations and the theoretical predictions for the growth rate of the convective layer in cases with constant driving flux, both with and without rotation (Equations~\ref{eq:h_FF_no_rot}--\ref{eq:h_FF_rot}). However, when the driving flux decays with time, the models in Equations~\eqref{eq:h_DF_no_rot} and \eqref{eq:h_DF_rot} tend to overpredict the extent of the convection zone. The simulations show that the convection zone eventually stalls in these declining flux cases, even without rotation. The size of the stalled convection zone is different in the cases with and without rotation, however. In terms of the size of the box, the convection zone size is $h\approx 0.61$ in the non-rotating case, whereas with rotation it is only $h\approx 0.38$, nearly half as deep, showing that rotation limits the extent even though the qualitative behavior is similar. 

Whereas the analytic model in Equation~\eqref{eqn:hypothesis} (from which Equations~\eqref{eq:h_DF_no_rot} and \eqref{eq:h_DF_rot} are derived) assumes a constant mixing efficiency $\gamma$, the simulations show that $\gamma$ decreases significantly in the dwindling-flux cases, leading to a stalled convection zone even in the absence of rotation. This reduction in $\gamma$ likely reflects a change in the physical mechanism of entrainment at low fluxes, where viscous dissipation may compete with or even overwhelm buoyant driving, although further work will be needed to clarify the origin of this behavior.

Our findings could have important implications for 1D evolution codes \citep[e.g.,][]{Vazan2018,Muller2020,Arevalo2025,Sur2024,Sur2025}. While these models account for a declining luminosity over time, if the initial entropy is high enough (a hot start), the outer convection zone can reach deep into the planet, sometimes nearly all the way through. In contrast, for low-entropy initial conditions (cold starts), the outer convection zone stalls. However, we emphasize that such models rely on mixing-length theory and lack a physically motivated prescription for turbulent entrainment across convective boundaries. Incorporating entrainment laws such as the ones in this work could significantly improve the predictions of 1D evolution models. A first step in this direction was taken by \citet{Scott2021} in the context of core convection in stars, where they showed that including an entrainment-based mixing scheme helps to reproduce key observational features, such as the mass dependence of the main-sequence width in the HR diagram. Similar improvements could be expected in planetary evolution models with compositional gradients, likely improving constraints on the mass and size of fuzzy cores.

Our results may also offer new insight into the energetics of core erosion. Previous estimates by \citet{Helled2022}, and more recently by \citet{Fuentes_et_al_2025} suggest that if the energy released during the planet's cooling history is available to drive mixing (a scenario they called the ``global story''), then even a compact or fuzzy core could be fully mixed into the outer convective envelope. In contrast, if mixing is restricted to the thermal energy available near the core-envelope interface (the  ``local story''), then the energy is insufficient to fully erode any core. These two scenarios, however, neglect the role of rotation in modulating convective mixing. Our findings suggest that rapid rotation, combined with a declining cooling flux, can substantially reduce the mixing efficiency of convection, potentially inhibiting core erosion even in the global story.

In light of current observations of Jupiter and Saturn, there is no consensus on the extent of the well-mixed convective region, which determines the size of the fuzzy core \citep[see Figure 2 in][]{Helled2022}. For example, the models of \cite{Militzer2022}, constructed from Jupiter's zonal gravity data, and those in \cite{Idini2022}, based on tidal interactions between Io and Jupiter, constrain the extent of the fuzzy core up to $0.7R$, leaving the outer $30\%$ of the radius well-mixed and vigorously convective. In contrast, the models of \cite{Nettelmann2021} truncate the core at $\sim 0.4R$ for both Jupiter and Saturn. Although the analytic models of \citet{Hindman2023} predict that the convection zone is confined to the outer $30\%$ of Jupiter and $20\%$ of Saturn, we stress that both those models and our simulations remain too limited to provide a reliable estimate of the convection zone's depth.

We encourage that future work should relax some of the simplifications adopted in this work. Our simulations neglected density stratification, as we adopted the Boussinesq approximation, which formally applies when the vertical extent of the fluid is small compared to a density scale height. In real planets, however, convection is driven by surface cooling, while the region of interest (the core) lies many scale heights below. Therefore, it is not clear whether the entrainment rates seen in Boussinesq models extend to more realistic, stratified simulations. Including density stratification could modify mixing through several ways. The energy required to dredge up heavy material from depth is different, and the nature of convective mixing itself changes: Boussinesq convection homogenizes density, while stratified convection restores an adiabatic gradient. Stratification also enhances asymmetries between upflows and downflows, altering the net kinetic energy flux available for mixing. Therefore, hydrodynamical models that go beyond Boussinesq would be of great interest.

In addition, our simulations were performed in a Cartesian box where gravity and rotation are aligned, which is appropriate near the poles. Real planets are spherical, so gravity decreases radially with depth and the rotation vector is aligned with gravity only at the poles. \citet{Fuentes2025} investigated semiconvective mixing in the full sphere, and showed that the latitude-dependent misalignment of gravity and rotation, combined with the radial variation of gravity, alters the buoyancy flux required for mixing. A non-constant gravity profile changes the potential energy cost of lifting heavy elements upward and therefore modifies the entrainment rate. They also found that these effects produce anisotropic mixing. In particular, Figures 6 and 7 of \citet{Fuentes2025} show stronger heat and composition fluxes near the equator than at higher latitudes, favoring cylindrical rather than spherical mixing and raising the possibility of a non-spherical fuzzy core. The global simulations on the sphere also permit the development of large-scale flows such as meridional circulation and zonal flows that are inhibited in Cartesian domains.

\begin{acknowledgements}
J.R.F. is supported by the Sherman Fairchild Postdoctoral Fellowship at Caltech, and NASA Solar System Workings grant 80NSSC24K0927. A.C. and S.Z. acknowledge support
from NSERC Discovery Grant RGPIN-2023-03620. S.Z. was supported by a Max Stern Recruitment Fellowship from McGill University.
\end{acknowledgements}

\appendix \label{sec:appendix}

\section{Fluid equations and numerical methods}

We express the fluid quantities as the sum of a linear hydrostatic background (denoted by the subscript 0) and a dynamic perturbation to the background (denoted by primes), e.g., the total temperature and composition are expressed as $T = T_0(z) + T'$, and $C = C_0(z) + C'$, respectively. The density perturbations satisfy $\rho'/\rho_0 \ll 1$, and are related to $T'$ and $C'$ through $\rho' = \rho_0(\beta C' - \alpha T')$, as demanded by the Boussinesq approximation \citep{Spiegel_Veronis_1960}. Here, $\beta$ and $\alpha$ are the coefficients of compositional and thermal contraction/expansion (both assumed positive constants), respectively.
Before presenting the fluid equations, we non-dimensionalize them using $[T] = |\partial_z T_0| H$, $[C] = |\partial_z C_0| H$ as units of temperature and composition. We use the domain's depth $H$ as the unit of length, and the thermal diffusion time $\tau_T = H^2/\kappa_T$ (where $\kappa_T$ is the thermal diffusivity) as the unit of time. By this choice, a unit of pressure corresponds to $[P] = \rho_0 (\kappa_T/H)^2$. The dimensionless equations are 

\begin{gather}
\nabla \cdot \bm{u} = 0\, , \label{eq:div u}\\
\nonumber \dfrac{\partial \bm{u}}{\partial t} + \bm{u}\cdot \nabla \bm{u} + \dfrac{\rm Pr}{\rm Ek} \bm{\hat{z}}\times \bm{u}  = - \nabla P + \mathrm{Ra}\mathrm{Pr}\left(R_\rho C - T\right)\bm{\hat{z}}\\
+ \mathrm{Pr} \nabla^2\bm{u} \, ,\\
\dfrac{\partial C}{\partial t} + \bm{u}\cdot \nabla C = \tau \nabla^2 C\, ,\\
\dfrac{\partial T}{\partial t} + \bm{u}\cdot \nabla T = \nabla^2 T\, , \label{eq:T}
\end{gather}
where $D/Dt = \partial_t + \bm{u}\cdot \bm{\nabla}$, and $\bm{u}$ is the velocity field.

There are 5 dimensionless numbers that characterize the evolution of the flow. These are the Rayleigh number Ra, Ekman number Ek, stability ratio $R_\rho$, Prandtl number Pr, and diffusivity ratio $\tau$, which are defined respectively as
\begin{align}
    &\mathrm{Ra} = \frac{\alpha g |\partial_z T_0|H^4}{\kappa_T \nu} ,\hspace{0.1cm}
    \mathrm{Ek} = \frac{\nu}{2\Omega H^2},\\
   & R_\rho = \frac{\beta |\partial_z C_0|}{\alpha |\partial_z T_0|},\hspace{0.1cm}
    \Pran = \frac{\nu}{\kappa_T},\hspace{0.1cm}
    \tau = \frac{\kappa_C}{\kappa_T},
\end{align}
where $\Omega$ is the rotation rate, $\nu$ is the kinematic viscosity, and $\kappa_C$ is the compositional diffusivity. $R_\rho$ is the ratio of the stabilizing and destabilizing effect of the compositional and thermal buoyancy.

We initialize the fluid with uniform temperature profile, $T_0(z) = 1$, and a linear composition gradient, $C_0(z) = 1-z$. This setup ensures that the initial stratification is stable against both convection and double-diffusive instabilities, regardless of the value of the input stability ratio $R_\rho$. We impose zero compositional flux at both the top and bottom boundaries, $\partial_z C|_{z=0,1} = 0$. Similarly, the thermal boundary conditions consist of zero heat flux at the bottom, $\partial_z T|_{z=0} =0$), and a time-dependent (or constant) heat flux at the top, $\partial_z T|_{z=1} = f(t)$, where $f(t) = -1$ for the cases of constant cooling flux, and $f(t) = -t_\star/(t+t_\star)$, with $t_\star \sim 5\times 10^{-3}$, for the cases with declining flux. The boundary conditions for the velocity are impenetrable and stress free at both boundaries ($\hat{z}\cdot\bm{u} = \hat{x}\cdot\partial_z\bm{u} = \hat{y}\cdot\partial_z\bm{u} = 0$ at $z = 0,1$).

All the simulations in this work use the following input parameters: Rayleigh number $\mathrm{Ra} = 2\times 10^{10}$, stability ratio $R_\rho = 0.1$, Prandtl number $\mathrm{Pr} = 0.1$, diffusivity ratio $\tau = 0.1$, and Ekman number $\mathrm{Ek} = 3 \times 10^{-6}$. For the non-rotating cases, we set the Coriolis force to zero by taking $\mathrm{Ek} = \infty$. In comparison, under Jovian interior conditions, typical values are $\mathrm{Ra} = \mathcal{O}(10^{43})$, $\mathrm{Ek} = \mathcal{O}(10^{-17})$, and $\tau \approx \mathrm{Pr} = \mathcal{O}(10^{-2})$ \citep[see, e.g.][]{Guillot_2004,French2012,Dhouib2024}. The value of the stability ratio $R_\rho$ remains uncertain due to limited knowledge of the thermal and compositional gradients deep inside gas giants, but we expect it to vary from $R_\rho \lesssim 1$ in convective regions, to $R_\rho \gg 1$ in the core. Nevertheless, by selecting parameters such that $\mathrm{Ra}\gg 1$, $\mathrm{Ek} \ll 1$, $\mathrm{Pr} < 1$, and $\mathrm{Le^{-1} < 1}$, we ensure that our simulations lie within the same qualitative dynamical regime as that of gas giant interiors.

We time-evolve Equations \eqref{eq:div u}--\eqref{eq:T} using the Dedalus pseudospectral solver \citep{Burns2020} version 3, using timestepper SBDF2 \citep{wang_ruuth_2008} and CFL safety factor of 0.2. All variables are represented using a Chebyshev series with 384 terms for $z \in [0, 1]$ and Fourier series with 384 terms in the periodic $x$ and $y$ directions, where $x,~y \in [0, 1]$. We use the 3/2-dealiasing rule in all directions, so that nonlinearities are calculated in physical space on a $576^3$ grid. To start our simulations, we add random-noise temperature perturbations sampled from a normal distribution with a magnitude of $10^{-5}$ compared to the initial temperature field.

\bibliographystyle{aasjournal}


\end{document}